\documentclass[aps,preprint,floats,epsf,epsfig,nofootinbib,letter]{revtex4}

\usepackage{epsfig}
\usepackage{graphicx}
\usepackage{dcolumn}
\usepackage{bm}
\usepackage{subfigure}

\def\DESepsf(#1 width #2){\epsfxsize=#2 \epsfbox{#1}}

\begin{document}
\def\be{\begin{eqnarray}}
\def\en{\end{eqnarray}}
\def\non{\nonumber}
\def\la{\langle}
\def\ra{\rangle}
\def\Br{{\mathcal B}}
\def\BB{{{\cal B} \overline {\cal B}}}
\def\BD{{{\cal B} \overline {\cal D}}}
\def\DB{{{\cal D} \overline {\cal B}}}
\def\DD{{{\cal D} \overline {\cal D}}}
\def\sq{\sqrt}


\title{Classical random walks over complex networks
and network complexity}

\author{Chih-Lung Chou}
\email{choucl@cycu.edu.tw}
\affiliation{Department of Physics,
Chung Yuan Christian University,
Taoyuan City 32023, Taiwan }

\date{\today}

\begin{abstract}
In this paper we view the steady states of classical random walks
over complex networks with an arbitrary degree distribution as
states in thermal equilibrium. By identifying the distribution of
states as a canonical ensemble, we are able to define the
temperature and the Hamiltonian for the random walk systems. We
then calculate the Helmholtz free energy, the average energy, and
the entropy for the thermal equilibrium states. The results shows
equipartition of energy for the average energy. The entropy is
found to consist of two parts. The first part decreases as the
number of walkers increases. The second part of the entropy
depends solely on the topology of the network, and will increase
when more edges or nodes are added to the network. We compare the
topological part of entropy with some of the network descriptors
and find that the topological entropy could be used as a measure
of network complexity. In addition, we discuss the scenario that a
walker has a prior probability of resting on the same node at the
next time step, and find that the effect of the prior resting
probabilities is equivalent to increasing the degree for every
node in the network.
\end{abstract}


\maketitle
\section{Introduction}

Complex networks describe a variety of systems of importance such
as the Internet, the World Wide Web, the cell, and social networks
\citep{general, Evolution, cell}. For example, the Internet is a
complex network that consists of billions of devices that are
connected by physical links \citep{internet}; Social networks are
made up of humans or organizations that are linked by various
social relations \citep{social}; The World Wide Web is a network
with large amounts of web pages being connected by hyperlinks.
Traditionally large scaled networks have been described as random
graphs by Erd\H{o}s-R\'{e}nyi model which have every pair of nodes
connected with probability $p$ such that edges are distributed
randomly \citep{renyi, renyi2}. In recent years the increased
computing power and the computerization of data acquisition allow
scientists to investigate real large scaled networks and find
deviations of the topologies of real large networks from a random
graph \citep{WWW, smallworld, richclub}. In most real networks the
clustering coefficient is much larger than it is in comparable
random graphs. The distribution of node degrees in many real
networks significantly deviated from a Poisson distribution that
is predicted by a random graph. In fact, the World Wide Web and
the Internet have a power-law tail in their degree distribution
thus are scale-free networks \citep{WWW, smallworld}. New models
and methods are thus needed to understand the topology, the
underlying organizing principles, and various dynamics that take
place on networks \citep{NetScaling, ScaleFreeRandom,
MergingCliques}. For example, statistical mechanics offers a
framework for describing the topology and the evolution of these
network systems, and also provides tools and measurements to
quantitatively depict these organizing principles
\citep{NetStatistics}.

Random walks provides an explanation for many stochastic processes
in various fields such as chemistry, computer science, physics,
and ecology \citep{randomwalk}. Usually, random walks are assumed
to be Markov processes \citep{Markov}. A random walk in discrete
time steps on a complex network is a special case of a Markov
chain and can be viewed as a generalization of Drunkard's walk
\citep{drunkard}. When a walker is on node $a$ of the network the
walker picks the available edges that are linked to the node with
equal probability. Thus, if node $a$ has $K_a$ edges the walker
will go to each one with probability $1/K_a$ at the next time
step. We can ask the questions like what is the average time for
the walker to return to its starting node? What is the average
time for the walker to reach another node $b$ on the network? A
quantity called the Mean First Passage Time (MFPT) gives answers
to the questions \citep{MFPT}. The MFPT $\langle T_{aa}\rangle $
for a walker on node $a$ to return to the same node is $K_a/K$,
depending only on the degree $K_a$ and the total number of degree
$K$ of the network. Studies on MFPT between two different nodes
$a$ and $b$ show asymmetry between $\langle T_{ab} \rangle $ and
$\langle T_{ba}\rangle $. Often the asymmetry between the MFPT has
explicit degree dependence. The results can be used to estimate
the size of large networks and provide a measure for effectiveness
in communication between nodes.

Random walks over complex networks also possess another intriguing
property. A random walk is ergodic and after a long time the
probability $P_a$ of finding a walker on node $a$ is solely
determined by the degree $K_a$ of the node and the total degree
$K$. The probability distribution of $P_a$ always reaches a steady
distribution, independent of the initial position for the walker
\citep{steadystate}. The process is similar to the processes in
thermal systems that go from non-equilibrium states to equilibrium
states. Independent of various initial conditions, a thermal
system could reach the same equilibrium state in certain
conditions after a relatively long time. For example, the free
expansion of ideal gas and the diffusion of ink drops in a glass
of water. It thus suggests a random walk on network to be viewed
as a thermal system. In classical thermodynamics the state of a
thermal system is defined as a condition uniquely specified by a
set of properties \citep{thermodynamics}. Here we use the
probability distribution $\{P_a\}$ to specify the states of a
random walk on network. The topological factors of a network such
as the number of nodes $M$, the degree distribution $\{K_a\}$, and
the distribution of edges linking the nodes are viewed as the
parameters that specify the phase space of a random walk system.
Therefore, an equilibrium state in the systems of random walk over
networks is one in which the probability distribution $\{P_a\}$
does not change with time unless the system is acted upon by
external influences. A non-equilibrium state has its probability
distribution $\{P_a(t)\}$ vary with time. A stochastic process in
the random walk system is thus a path that consists of a series of
states through which the system passes. In statistical
thermodynamics a thermodynamic system is regarded as an assembly
of enormous number of ever-changing microstates. The basic
assumption of statistical thermodynamics is that all microstates
of an assembly are equally probable. The thermal equilibrium state
is defined as the most probable state that has the largest number
of corresponding microstates in a thermodynamic system. In section
\ref{sec:canonical} we find the probability of the distribution of
$N$ non-interacting random walkers $\{n_a\}$ that gives the steady
state as the most probable state in the thermodynamic system.

In this paper we consider an arbitrary finite network which
consists of nodes $a=1,2,\dots,M$ with undirected edges. The
network is assumed to be connected, i.e, there is at least a path
between each pair of nodes ($a,b$). The connectivity of the
network is represented by the off-diagonal adjacency matrix $A$
with its elements $A_{ab}$ to be either $1$ (if $(a,b)$ are
directly connected) or $0$ (if $(a,b)$ are not directly
connected). The degree $K_a$ of node $a$ is defined to be the
total number of edges that link directly to node $a$ from other
nodes in the network. A classical walker moving on the network is
stochastic and can be described by the master equation
\begin{equation}
|P(t+1)\rangle = AD^{-1}|P(t)\rangle. \label{masterEq}
\end{equation}
\noindent Here $A$ is the adjacency matrix, $D$ denotes a $M\times
M$ diagonal matrix with diagonal elements $D_{aa}=K_a$, and
$|P(t)\rangle$ is a $M\times 1$ column vector whose element
$P_a(t)$ is the probability of finding a walker on node $a$ at
time step $t$. Independent of the initial conditions
$|P(t)\rangle$ is shown to approach to a steady probability
distribution $|\pi \rangle$ at large time $t \to \infty$ with
\begin{equation}
\pi_a = \frac{K_a}{K},\label{steadyState}
\end{equation}
\noindent where $\pi_a$ is the element of $|\pi\rangle$, and
$K\equiv \sum_a K_a$ is the sum of all degrees. From the result in
(\ref{steadyState}) the average return time $\langle
T_{aa}\rangle$ is easily derived as follows. Consider a walker
that moves on the network from time step $t_0$ to $t_0+\Delta t$
($\Delta t\gg 1$). Initially at time step $t_0$ the walker is on
node $a$. During the time interval the expected number of finding
the walker on node $a$ is $\pi_a\Delta t$, and thus the average
return time for node $a$ is found by $\langle T_{aa}\rangle =
\Delta t/(\pi_a \Delta t) = K/K_a$. Other interesting quantities
about the network such as the mean first passage time $\langle
T_{ab}\rangle$ and the random walk centrality can also be found by
studying the master equation in (\ref{masterEq}) \citep{MFPT}.

Although the master equation provides detailed information about
the motion of a walker on the network, the fact that the
probability distribution $|P(t)\rangle$ finally approaches to a
steady state $|\pi\rangle$ does suggest that $|\pi\rangle$ can be
thought of as a thermal equilibrium state in a thermal system. For
a random walk on network with $N$ non-interacting walkers ($N \gg
1$), the system should have a ``temperature" $T$ when it reaches
its thermal equilibrium. Thus $(N, M, K, \{K_a\}, T)$  are viewed
as state variables for the random walk thermal system. In section
\ref{sec:steadystate} we use the result in (\ref{steadyState}) to
help us define the so-called ``temperature" $T$ for the steady
states of classical random walks on networks. We then calculate
various state functions such as the internal energy, the entropy,
and the Helmholtz free energy for the steady state (viewed as a
thermal equilibrium state) in section \ref{sec:canonical}. We find
that the topological part of the entropy can be used as a measure
of network complexity. In section \ref{sec:RestingProb} we discuss
a modified random walk model that has prior resting probabilities
for a walker to stay on nodes at the next time step. In the last
section we compare the topological entropy to other network
descriptors and give our conclusion.

\section{Steady states as thermal equilibrium states\label{sec:steadystate}}
\begin{figure}[t!]
\includegraphics[width=0.6\textwidth,natwidth=610,natheight=642]{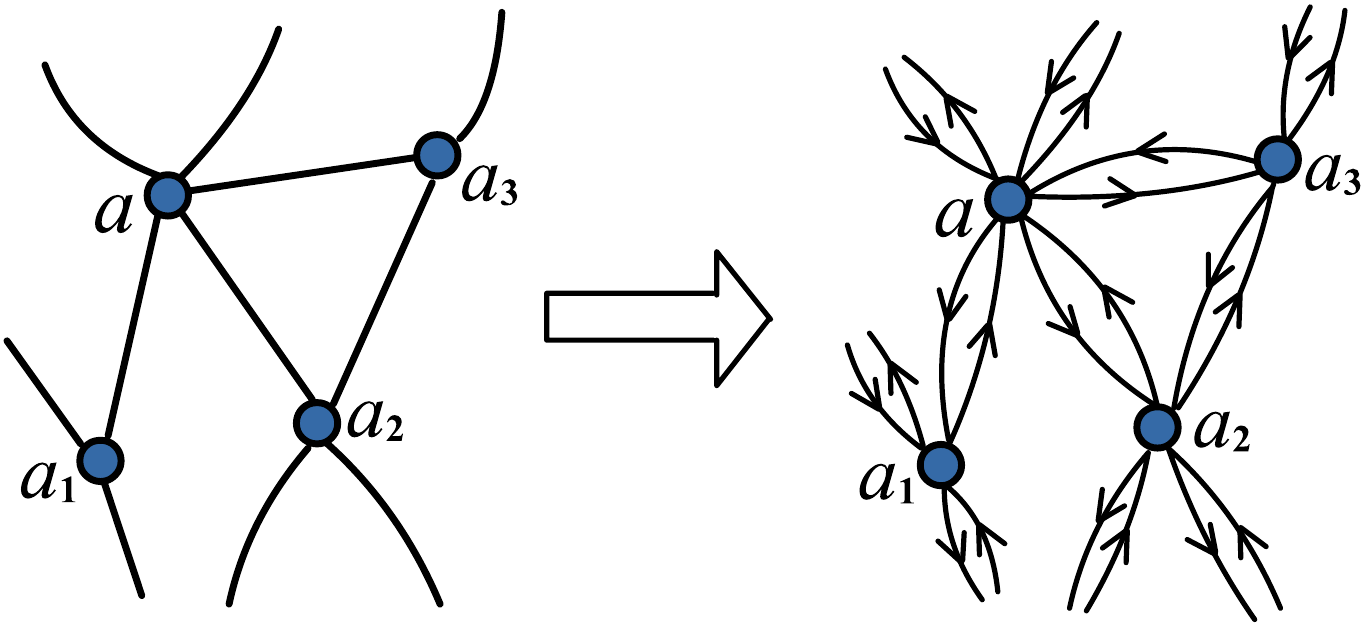}
\caption{Any undirected edge in the network can be viewed as
consisting of an outgoing edge and an incoming edge in the random
walk models. }
\label{fig1}
\end{figure}

Instead of considering a single walker on a network, we assume $N$
($N \gg 1$) mutually non-interacting classical walkers that move
on a connected complex network with $M$ nodes. The move of walkers
is ergodic and will reach an equilibrium after a substantial long
period of time. It strongly suggests that we view the steady state
as a thermal state in equilibrium. As shown in Fig.\ref{fig1}, any
undirected edge of the network can be viewed as consisting of two
directed links. Each node in the network has equal numbers of
outgoing and incoming links attached to it. In the steady state,
the distribution of walkers $\{n_a\}$ does not change in time.
This is only possible when all the directed links, outgoing and
incoming, have the same flux of walkers at each time step. It is
easily checked that the steady state has the walker distribution
$\{n_a =NK_a/K\}$ or equivalently, the probability distribution
$\{P_a = K_a/K\}$.

\begin{figure}
\includegraphics[width=0.6\textwidth,natwidth=610,natheight=642]{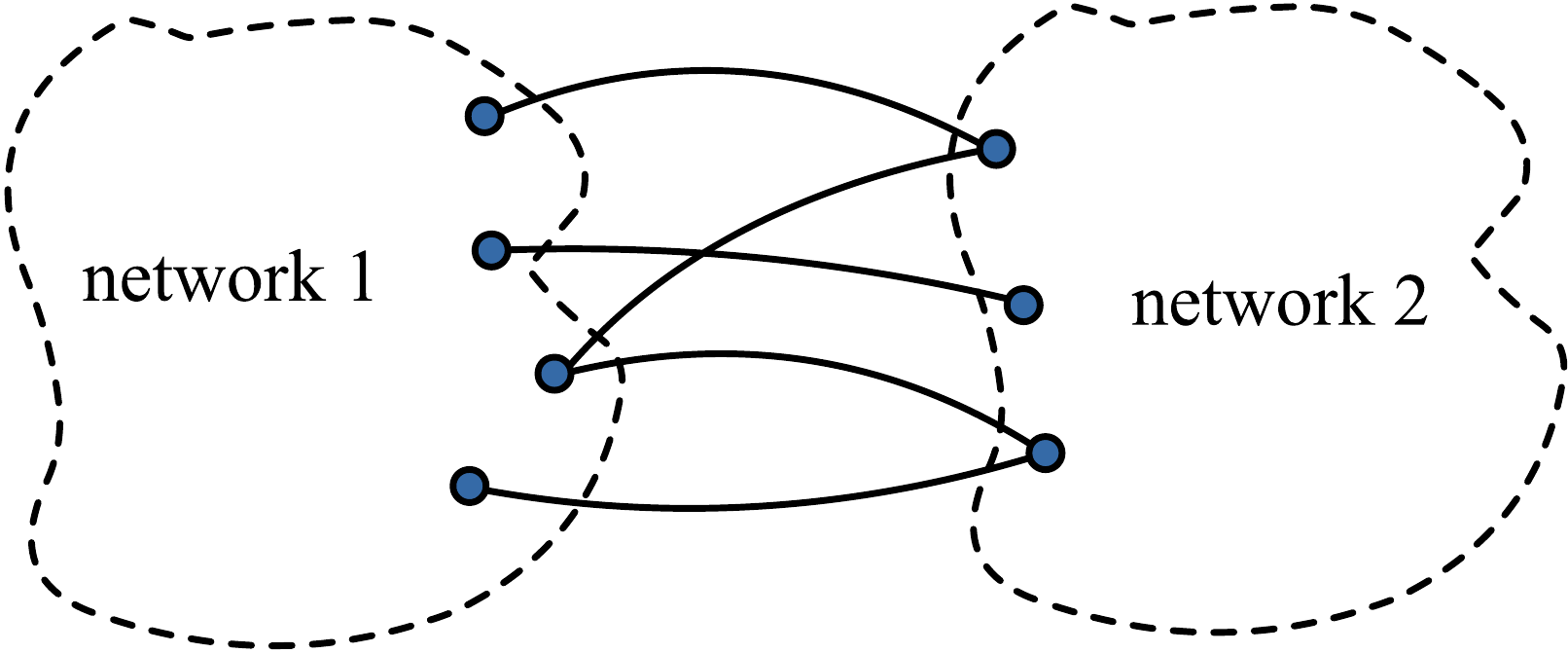}
\caption{The network $1$ with $M_1$ nodes and total degree
$K^{(1)}$ is linked to the network $2$ with $M_2$ nodes and total
degree $K^{(2)}$. The number of linking edges $\ell$ between the
network $1$ and $2$ is assumed to be small such that $\ell <<
K^{(1)}, K^{(2)}$ is satisfied. In the figure only the linked
nodes between the two networks are shown.}\label{fig2}
\end{figure}

In classical thermodynamics the concept of temperature is a
property of a system that determines if thermal equilibrium exists
with some other system. Once the steady state of random walks is
viewed as a thermal equilibrium state then it is possible to
define the ``temperature" for the steady state. To find the
temperature let's consider the contact of two random walk systems,
the network $1$ of $N_1$ walkers with $M_1$ nodes and the network
$2$ of $N_2$ walkers with $M_2$ nodes. Both random walk systems
are assumed to be in their steady states. So the number of walkers
on node $a$ in the network $1$ is $n^{(1)}_a=N_1
K^{(1)}_a/K^{(1)}$, where $K^{(1)}_a$ is the degree of node $a$
and $K^{(1)}$ is the total degree of the network $1$. Similarly,
the number of walkers on node $b$ in the network $2$ is
$n^{(2)}_b=N_2 K^{(2)}_b/K^{(2)}$ with $K^{(2)}_b$ being the
degree of node $b$ and $K^{(2)}$ being the total degree of the
network $2$. We then bring the two systems together and join them
by linking $\Delta M_1$ nodes of the network $1$ to $\Delta M_2$
nodes of the network $2$ as shown in Figure \ref{fig2}. The number
of linking edges $\ell$ between the two networks is assumed to be
much smaller than $K^{(1)}$ and $K^{(2)}$. Under the small $\ell$
assumption the total degree of the combined network can be thought
of as the sum of the degrees of the two networks. The combined
network now has $N=N_1+N_2$ walkers moving on it with $M=M_1+M_2$
nodes and the total degree $K=K^{(1)}+K^{(2)}$. When the
distribution of walkers reaches an equilibrium in the combined
system , the number of walkers $\tilde n^{(1)}_a$ on node $a$ in
the network $1$ and the the number of walkers $\tilde n^{(2)}_b$
on node $b$ in the network $2$ are
\begin{eqnarray}
\tilde n^{(1)}_a = \frac{N_1+N_2}{K^{(1)}+K^{(2)}} K^{(1)}_a, \label{newn1a}\\
\tilde n^{(2)}_b = \frac{N_1+N_2}{K^{(1)}+K^{(2)}} K^{(2)}_b.
\label{newn2b}
\end{eqnarray}
\noindent The differences $(\tilde n^{(1)}_a-n^{(1)}_a)$ and
$(\tilde n^{(2)}_b-n^{(2)}_b)$ are thus found as
\begin{eqnarray}
\tilde n^{(1)}_a-n^{(1)}_a = \frac{N_1N_2}{K^{(1)}K}(\frac{K^{(1)}}{N_1}-\frac{K^{(2)}}{N_2}) K^{(1)}_a, \label{diff1a} \\
\tilde n^{(2)}_b-n^{(2)}_b
=\frac{N_1N_2}{K^{(2)}K}(\frac{K^{(2)}}{N_2}-\frac{K^{(1)}}{N_1})K^{(2)}_b.\label{diff2b}
\end{eqnarray}
\noindent From the results in Eq.s(\ref{diff1a},\ref{diff2b}) the
net flow of walkers will go from network $2$ to network $1$ if
$K^{(1)}/N_1> K^{(2)}/N_2$ is satisfied. When $K^{(1)}/N_1<
K^{(2)}/N_2$ the net flow of walkers will go from network $1$ to
network $2$. There is no net flow of walkers between the two
networks if the ratio $K^{(1)}/N_1$ equals $K^{(2)}/N_2$. It thus
suggests that the ratio $K/N$ could be the candidate of
``temperature" for the steady state of the random walk on a
network of $N$ walkers with total degree $K$. It is interesting to
note that, with the ratio $K/N$ being identified as the
temperature for a random walk in equilibrium, the final
temperature after combining two equilibrium random walks is always
between the two temperatures of the two random walks before
combining them. In the next section more reasons will be provided
for identifying $K/N$ as the temperature.

\section{Canonical distribution for classical random walks on networks\label{sec:canonical}}
The probability of obtaining the distribution of walkers $\{n_a\}$
over nodes is found as
\begin{equation}
P(\{n_a\};N,\{K_a\}) = \frac{N!}{K^N}\prod_{a}
\frac{K_a^{n_a}}{n_a !}. \label{walkerDistribution}
\end{equation}
\noindent The result in (\ref{walkerDistribution}) is obtained by
using the probability distribution $\{P_a = K_a/K\}$ for a single
walker and then calculate the probability of having the walker
distribution $\{n_a\}$. \noindent In the large $N$ limit, it has
\begin{eqnarray}
\ln P(\{n_a\};N,\{K_a\}) =-\frac{N}{K}\sum_a \xi_a
\ln(\frac{\xi_a}{K_a})+O(\ln N).\label{lnP}
\end{eqnarray}
\noindent Here $\xi_a \equiv n_a K/N$ ranges from $0$ to $K$. The
steady state $|\pi\rangle$ corresponds to the distribution of
$\xi_a$, $\{\xi_a = K_a\}$. From (\ref{lnP}), we find the relative
frequency $\Omega$ for the distribution $\{\xi_a\}$
\begin{eqnarray}
\Omega(\{\xi_a\}, \frac{K}{N}, \{K_a\}) =
\exp[{-\frac{N}{K}H(\{\xi_a\}; \{K_a\})}], \label{Omega}
\end{eqnarray}
\noindent with
\begin{eqnarray}
 H(\{\xi_a\};\{K_a\}) \equiv \sum_a \xi_a \ln(\frac{\xi_a}{K_a}).\label{Hamiltonian}
\end{eqnarray}
\noindent The relative frequency $\Omega$ resembles the canonical
distribution for a thermal system with Hamiltonian $H(\{\xi_a\};
\{K_a\})$ that is in equilibrium with a heat bath at temperature
$T=K/N$. The Hamiltonian $H(\{\xi_a\}; \{K_a\})$ is easily proved
to be non-negative and has the minimum $H_{min}=0$ which
corresponds to the steady state, i.e., $\{\xi_a = K_a\}$. The
temperature is identified as $T=K/N$ since when $\beta =1/T$
increases the expectation value of $\langle H \rangle$ will
decrease and the system is likely to lie in the low-energy states.

\subsection{Partition function, Helmholtz free energy, and the entropy}
The partition function for the random walk system is calculated as
follows. Consider the expansion of Hamiltonian
$H(\{\xi_a\};\{K_a\})$
\begin{equation} H(\{\xi_a\}; \{K_a\}) =
\sum_{a=1}^M \frac{(\xi_a-K_a)^2}{2K_a} + O(\Delta \xi^3).
\label{HExpansion}
\end{equation}
\noindent Here $\Delta \xi$ and $(\xi_a-K_a)$ are of the same
order of magnitude. The partition function $Q_M$ for the thermal
system is defined as
\begin{equation}
Q_M \equiv \sum_{\{\xi_a\}} \exp[-\beta H(\{\xi_a\}; \{K_a\})].
\label{QM}
\end{equation}
\noindent Define new variables $\chi_a \equiv (\xi_a-K_a)$ so that
$\sum_a \chi_a =0$ is satisfied. We then calculate $Q_M$ and get
\begin{eqnarray}
Q_M &\simeq& \int \{\prod_{a=1}^M d\chi_a\} \delta(\sum_a
\chi_a)\exp[-\sum_{a=1}^M \frac{\beta \chi_a^2}{2K_a}] \nonumber \\
&=&(\frac{2\pi}{\beta})^{(M-1)/2} K^{-1/2}\prod_{a=1}^M K_a^{1/2}.
\label{QM2}
\end{eqnarray}
\noindent From the partition function $Q_M$ in (\ref{QM2}), we
find the average ``energy" for the random walk system in
equilibrium
\begin{equation}
\langle H \rangle = -\frac{\partial \ln Q_M}{\partial \beta} =
\frac{(M-1)K}{2N} = \frac{(M-1)}{2}T. \label{aveH}
\end{equation}
\noindent The result in (\ref{aveH}) is exactly the equipartition
of energy in thermodynamics. The total degree of freedom for the
random walk system is $(M-1)$ as can be easily seen from the
number of independent variables in the Hamiltonian $H(\{\xi_a\};
\{K_a\})$, and each degree of freedom acquires the same average
energy $T/2$.

The Helmholtz free energy $F$ for the random walk system is found
by
\begin{equation}
F\equiv \frac{-1}{\beta}\ln Q_m = -\frac{T}{2}\{(M-1)\ln(2\pi
T)-\ln K+ \sum_{a=1}^M \ln K_a  \}. \label{FreeEnergy}
\end{equation}
\noindent From the Helmholtz free energy we obtain the entropy $S$
for the system
\begin{equation}
S\equiv -\frac{\partial F}{\partial T}=\frac{M-1}{2}[\ln(2\pi
T)+1] -\frac{1}{2}\ln K +\frac{1}{2} \sum_{a=1}^M \ln
K_a.\label{entropy}
\end{equation}
\noindent Obviously, the entropy $S$ in (\ref{entropy}) does
satisfy the relation $\partial S/\partial \langle H\rangle = 1/T$.
All state functions in the equations
(\ref{aveH}$\sim$\ref{entropy}) depend (at least partially) on the
state variables $M, T, K$, and $\{K_a\}$.

The entropy $S$ consists of two parts
\begin{eqnarray}
S&=&-\frac{M-1}{2}\ln N + S_{top}, \label{StwoParts}\\
S_{top}&=&\frac{M-1}{2}\ln (2\pi e) + \frac{M-2}{2}\ln K
+\frac{1}{2}\sum_{a=1}^M \ln K_a, \label{Stopology}
\end{eqnarray}
\noindent where $S_{top}$ denotes the part solely determined by
the topology of the complex network. In general, without changing
the network topology the entropy $S$ decreases as the total number
of walkers $N$ increases. On the other hand $S_{top}$ seems like
to be an interesting property of the complex network. First, the
topological entropy $S_{top}$ will increase by adding more edges
to the network. For example, an edge is added to the network
between node $a$ and node $b$. As a result, both the degree $K_a$
and $K_b$ are increased by $1$, the total degree $K$ is increased
by $2$, and the topological entropy is changed by $\Delta S_{top}$
\begin{equation}
\Delta S_{top} =
\frac{M-2}{2}\ln(1+\frac{2}{K})+\frac{1}{2}\{\ln(1+\frac{1}{K_a})+\ln(1+\frac{1}{K_b})
\} >0. \label{DeltaStop}
\end{equation}
\noindent Second, adding more nodes to the network also increases
$S_{top}$. Suppose a new node is connected to nodes $1,2,\dots,q$
of the network. In the situation the number of nodes is increased
by $1$, all the degrees of nodes $1,2,\dots,q$ are increased by
$1$, and the total degree is increased by $2q$. Thus the change in
$S_{top}$ is
\begin{eqnarray}
\Delta S_{top}
&=&\frac{M-1}{2}\ln(1+\frac{2q}{K})+\frac{1}{2}[\ln(2\pi K)+1]
+\frac{1}{2}\sum_{a=1}^q\ln(1+\frac{1}{K_a}) +\frac{1}{2}\ln q
\nonumber \\
&>& 0. \label{DeltaStop2}
\end{eqnarray}
\noindent From the results in (\ref{DeltaStop}) and
(\ref{DeltaStop2}) we conclude that the topological entropy
$S_{top}$ always increases as more nodes or edges are added to
complex networks. It strongly suggests that the topological
entropy $S_{top}$ could be used as a measure of network
complexity. Among all connected networks with $M$ nodes, the star
topology (e.g. the network A in Fig.\ref{fig3} for $M=5$) has the
minimum value of $S_{top}$
\begin{equation}
S^{(min)}_{top}(M) = \frac{M-1}{2}\{\ln[2\pi(M-1)]+1 \}
+\frac{M-2}{2}\ln2, \label{StopMin}
\end{equation}
\noindent while the fully connected topology (e.g. the network L
in Fig.\ref{fig3} for $M=5$) has the maximum value of $S_{top}$
\begin{equation}
S^{(max)}_{top}(M) =\frac{M-1}{2}\{\ln[2\pi(M-1)^2]+1 \}
+\frac{M-2}{2}\ln M. \label{StopMax}
\end{equation}
\noindent As an example, let's consider four networks A, F, K and
L as shown in Figure \ref{fig3}. All the four networks have five
nodes but are of different topologies. Among the four networks the
star topology (network A) has the minimum value of $S_{top}$ while
the fully connected topology (network L) has the maximum
$S_{top}$. All nodes in the network A have just one edge linked to
them except for a central node that is connected to all other
nodes. Contrarily, the network L has no central node. Any two
nodes are linked by an edge in the network L. It has
$S^L_{top}=13.635>S^K_{top}=12.725>S^F_{top}=10.863>S^A_{top}=9.488$.
The above result suggests that the network A is less complex than
the network F, the network F is less complex than the network K,
and the network K is less complex than the network L.

\section{Random walk with a probability of resting on nodes }\label{sec:RestingProb}
In the section we consider a slightly different scenario: a walker
has a prior probability $\sigma_a$ of resting on node $a$ at each
time step. The scenario is indeed related to the random walks on
the connected networks that have self-linked edges linking a node
to itself. As an example let's consider a connected network with
self-linked edges as shown in Figure \ref{fig4}. A self-linked
edge is attached to node $a$ and two self-linked edges are
attached to node $c$. Since each self-linked edge contributes the
amount of $2$ to the degree, we thus find the degrees of the nodes
to be $\{K_a=4, K_b=2, K_c=6, K_d=2\}$ and the total degree of the
network is $K=14$. For a walker on node $c$ at time step $t$, it
has the probability $1/6$ of hopping to node $b$ or $d$ at the
next time step. In other words, it has the probability $2/3$ of
resting on node $c$ at time step $t+1$. Similarly, a walker on
node $a$ at time step $t$ could stay on the same node with the
probability $2/4$ at the next time step. From the distribution of
degrees we thus find the probabilities of finding a walker on the
nodes in the steady state, $\{P_a=4/14, P_b=2/14, P_c=6/14,
P_d=2/14\}$.

The discussion is easily generalized to the case of $N$
non-interacting random walkers moving on a connected network of
$M$ nodes with the degree distribution $\{K_a\}$ and the
distribution of prior resting probability on nodes $\{\sigma_a\}$.
Once again, here $\sigma_a$ denotes the probability of resting on
node $a$ at the next time step for a single walker. The case is
equivalent to a conventional random walk on a network of $M$ nodes
with the degree distribution $\{K_a^{\prime}\}$,
\begin{eqnarray}
K_a^{\prime} &=& K_a+2L_a = \frac{K_a}{(1-\sigma_a)}, \label{KaPrime}\\
 L_a &=& \frac{\sigma_a}{2(1-\sigma_a)}K_a,\label{La}
\end{eqnarray}
\noindent where $L_a$ denotes the number of the self-linked edges
attached to node $a$. In general, $L_a$ ranges from zero to
infinity and may not be an integer. We thus find the temperature
$T$ of the modified random walk model as
\begin{equation}
T= \frac{K^{\prime}}{N}, \label{T4modified}
\end{equation}
\noindent with the effective total degree $K^\prime$
\begin{equation}
K^{\prime} = \sum_a \frac{K_a}{(1-\sigma_a)}. \label{Kprime}
\end{equation}
\noindent The probability of having the walker distribution
$\{n_a\}$ is thus found by
\begin{equation}
P(\{n_a\};N, \{K_a\}, \{\sigma_a\}) = \frac{N!}{K^{\prime
N}}\prod_a \frac{1}{n_a !}[\frac{K_a}{(1-\sigma_a)}]^{n_a}.
\label{PnaModified}
\end{equation}
\noindent The probability distribution $P(\{n_a\};N, \{K_a\},
\{\sigma_a\})$ depends not only on the topology of the network but
also on the distribution of the node-resting probability
$\{\sigma_a\}$. In fact, it is better to view $\{\sigma_a\}$ as
one of the topological factors of the network. Other state
functions such as the average energy, the Helmholtz free energy,
and the entropy of the modified random walk system are obtained by
replacing $K_a$ with $K_a^{\prime}$, and replacing $K$ with
$K^\prime$ in the equations (\ref{aveH}, \ref{FreeEnergy},
\ref{entropy}). For example, the topological part of the entropy
in the modified random walk model is
\begin{equation}
S_{top}=\frac{M-1}{2}\ln (2\pi e) +\frac{M-2}{2}\ln K^\prime
+\frac{1}{2}\sum_{a=1}^M \ln K_a^{\prime}. \label{Stopology2}
\end{equation}
\noindent In conclusion, the effect of resting on nodes is
equivalent to increasing the degree at every node in the network.
The topological entropy also increases in the modified random walk
as compared to that in the conventional random walks without the
prior resting probability.

\section{Discussion}
In this paper we consider $N$ non-interacting classical walkers
that move on a connected network with $M$ nodes and the
distribution of degrees $\{K_a\}$. We found that the steady state
of the random walk system can be interpreted as a thermal state in
equilibrium.  We then identified the temperature for the system to
be $K/N$, where $K$ is the total degree, and calculated various
state functions of the system such as the average energy $\langle
H \rangle$, the Helmholtz free energy, and the entropy $S$. The
results show that the equilibrium thermal state has the property
of equipartition of energy. Without changing the network topology
the entropy $S$ increases as the total number of walkers
decreases. In fact, the entropy $S$ consists of two parts. The
first part depends solely on the number of walkers $N$ and the
number of nodes $M$. The second part, called the topological
entropy $S_{top}$ in the paper, depends solely on the network
topology. In general, $S_{top}$ will increase as more edges or
nodes are added to the network. It thus suggests that $S_{top}$
can be used as a measure of network complexity.

\begin{figure}
\includegraphics[width=0.5\textwidth,natwidth=610,natheight=642]{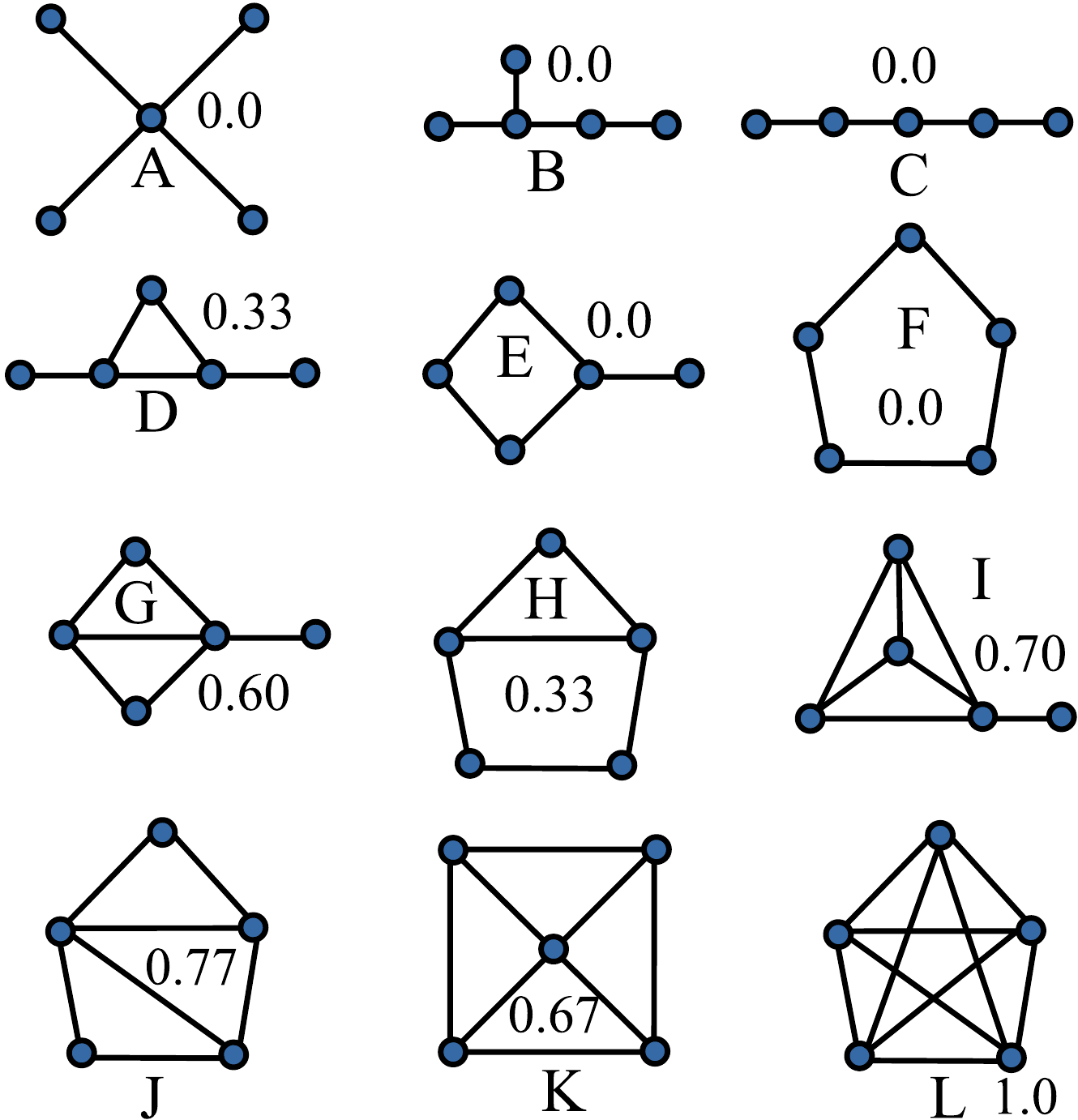}
\caption{The topological entropies for the five-node networks are
$S^A_{top}=9.488$, $S^B_{top}=9.691$, $S^C_{top}=9.835$,
$S^D_{top}=10.575$, $S^E_{top}=10.719$, $S^F_{top}=10.863$,
$S^G_{top}=11.339$, $S^H_{top}=11.541$, $S^I_{top}=11.975$,
$S^J_{top}=12.119$, $S^K_{top}=12.725$, and $S^L_{top}=13.635$.
The numbers in the figure denote the average clustering
coefficient $\langle C\rangle$ for each network.}\label{fig3}
\end{figure}

\begin{figure}
\includegraphics[width=0.33\textwidth,natwidth=610,natheight=642]{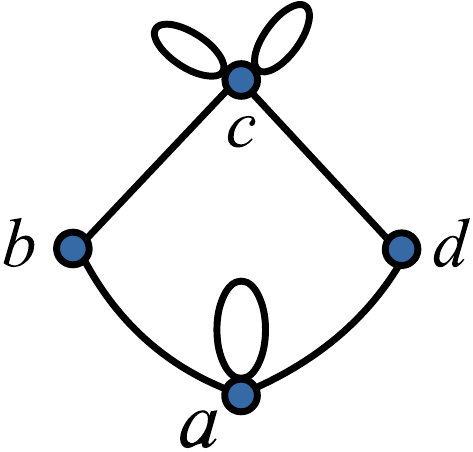}
\caption{A network with self-linked edges. Node $a$ has one
self-linked edge and node $c$ has two self-linked
edges.}\label{fig4}
\end{figure}

\begin{figure}
\includegraphics[width=0.6\textwidth,natwidth=610,natheight=642]{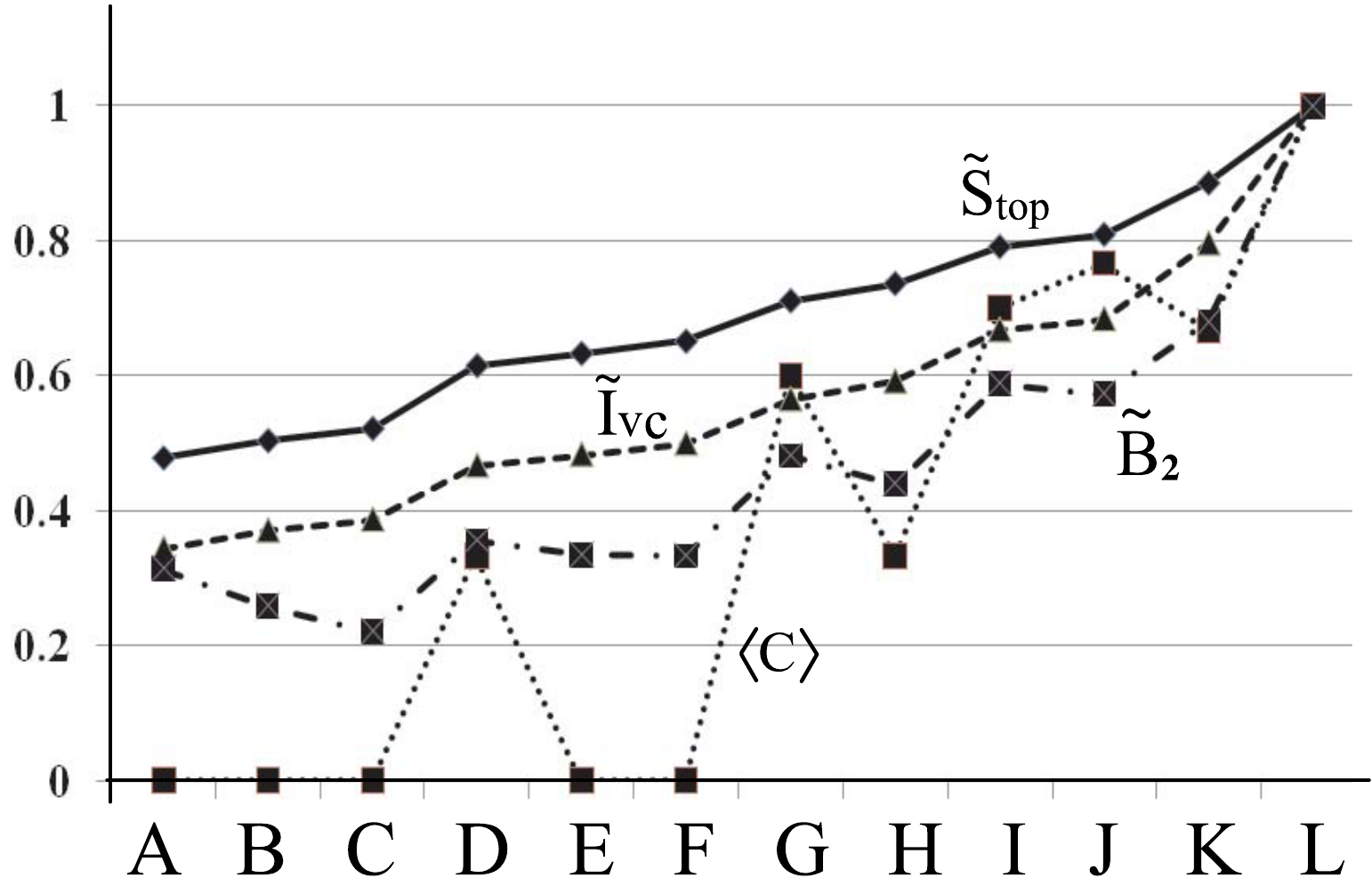}
\caption{Plots of network descriptors $\tilde S_{top}$, $\tilde
I_{vc}$, $\tilde B2$, and $\langle C \rangle$ for the five-vertex
graphs listed in Fig. \ref{fig3}. The descriptors $\tilde
S_{top}$, $\tilde I_{vc}$, and $\tilde B2$ are the normalized
indices of $S_{top}$, $I_{vc}$, and $B2$.}\label{fig5}
\end{figure}

There is no absolute definition of what complexity of networks
means. Even without a precise definition of network complexity, a
few descriptors are used as measures of network structure
\citep{complexity, descriptors}. For example, the information
theoretic index for node degree distribution
\begin{equation}
I_{vc}\equiv K \log_2 K-\sum_{a=1}^M K_a \log_2 K_a, \label{Ivc}
\end{equation}
\noindent which makes use of Shannon's formula for the total
information content of the vertex distribution. Usually $I_{vc}$
increases with the connectivity and other complexity factor such
as the number of cycles. As shown in Fig. \ref{fig3}, the total
degree $K$ increases from $K=8$ to $K=20$ for the five-node graphs
from A to L. The plot of the normalized index
\begin{equation}
\tilde I_{vc}= \frac{I_{vc}}{I^{(max)}_{vc}}, \label{normalIvc}
\end{equation}
is shown in Fig. \ref{fig5} with $I^{(max)}_{vc}$ being the
maximum of $I_{vc}$ for the five-node networks. It shows that
$I_{vc}$ does increase with connectivity. The number of cycles is
also considered as a stronger network complexity factor, and this
is seen in the sequence of graphs with one to five cycles: F
$\rightarrow$ H $\rightarrow$ J $\rightarrow$ K $\rightarrow$ L.
Another useful descriptor that describes the overall degree of
network clustering is the average clustering coefficient $\langle
C\rangle$
\begin{eqnarray}
\langle C \rangle &=& \frac{\sum_a C_a}{M}, \label{averageC} \\
C_a &=& \frac{2f_a}{K_a(K_a-1)}.
\end{eqnarray}
\noindent Here $f_a$ denotes the number of edges that link the
first neighbors of node $a$, and $C_a$ is the clustering
coefficient of node $a$. As seen in Fig. \ref{fig3}, the
clustering coefficients of all nodes in the acyclic graphs A, B,
and C, are zero. All vertices in cyclic graphs having four or more
vertices also have zero clustering coefficients. Nonzero
clustering coefficients can only be obtained in tri-membered
cycles. More tri-membered cycles in a graph usually give a higher
$\langle C\rangle $ value, as can be seen in the sequence of
graphs with one to three tri-membered cycles: D $\rightarrow$ G
$\rightarrow$ J. Other descriptors such as the $B2$ and $B3$
indices make use of vertex and vertex distance distribution
\begin{eqnarray}
B2&=& \sum_{a=1}^M \frac{K_a}{D_a}, \label{B2} \\
B3 &=& B2\log_2 B2 - \sum_{a=1}^M \frac{K_a}{D_a} \log_2
\frac{K_a}{D_a}. \label{B3}
\end{eqnarray}
\noindent Here $D_a$ is the sum of all the minimum distances
$D_{ab}$ between node $a$ and other nodes (node $b$)
\begin{equation}
D_a = \sum_{b \neq a} D_{ab}. \label{Distance}
\end{equation}
\noindent The indices $B2$ and $B3$ increase with the connectivity
and the number of cycles, similar to the index $I_{vc}$. However,
unlike the index $I_{vc}$, $B2$ also increases with the appearance
of an additional branch. This is seen in the plot of the
normalized $B2$ index in Fig. \ref{fig5} for the sequences of
graphs of the same total degree: C $\rightarrow$ B $\rightarrow$
A, F $\rightarrow$ E $\rightarrow$ D, H $\rightarrow$ G, J
$\rightarrow$ I.

In order to compare the topological entropy $S_{top}$ and the
other network descriptors $I_{vc}, \langle C\rangle$ and $B2$, the
descriptors are normalized and thus range from $0$ to $1$. For
example, the graph L in Fig. \ref{fig3} has the maximum value of
$I_{vc} = 46.439$ among all the five-vertex graphs. We thus define
the normalized index $\tilde I_{vc} = I_{vc}/46.439$. Similarly,
the normalized $B2$ index is defined as $\tilde B_2 = B_2/5.0$ for
all the five-vertex graphs in Fig. \ref{fig3}. The normalized
topological entropy $\tilde S_{top}$ for all $M$-node networks is
defined as
\begin{equation}
\tilde S_{top} = \frac{S_{top}-\frac{M-1}{2}\ln (2\pi
e)}{\{S_{top}-\frac{M-1}{2}\ln (2\pi e)\}_{max}}.
\label{StopNormal}
\end{equation}
\noindent The term $(M-1)\ln (2\pi e)/2$ is subtracted from
$S_{top}$ in equation (\ref{StopNormal}) since it is the same for
all $M$-node connected networks. A plot of $\tilde S_{top}$,
$\tilde I_{vc}$, $\langle C\rangle$, and $\tilde B_2$ for all the
five-node networks in Fig. \ref{fig3} is given in Fig. \ref{fig5}.
As easily seen from Fig. \ref{fig5}, $\tilde S_{top}$ and
$\tilde{I}_{vc}$ behave similarly as descriptors of network
structure. Both of them increase with the connectivity and the
number of cycles. Contrary to the $\tilde{B}2$ index,
$\tilde{S}_{top}$ will decrease with the appearance of an
additional branch. Thus the network with star topology will have
the minimum value of $S_{top}$, while the network of a chain of
nodes will have the minimum value of $B2$. All the descriptors in
Fig. \ref{fig5} will reach their maximum values when the network
has a fully connected topology.

\begin{figure}
\includegraphics[width=0.6\textwidth,natwidth=610,natheight=642]{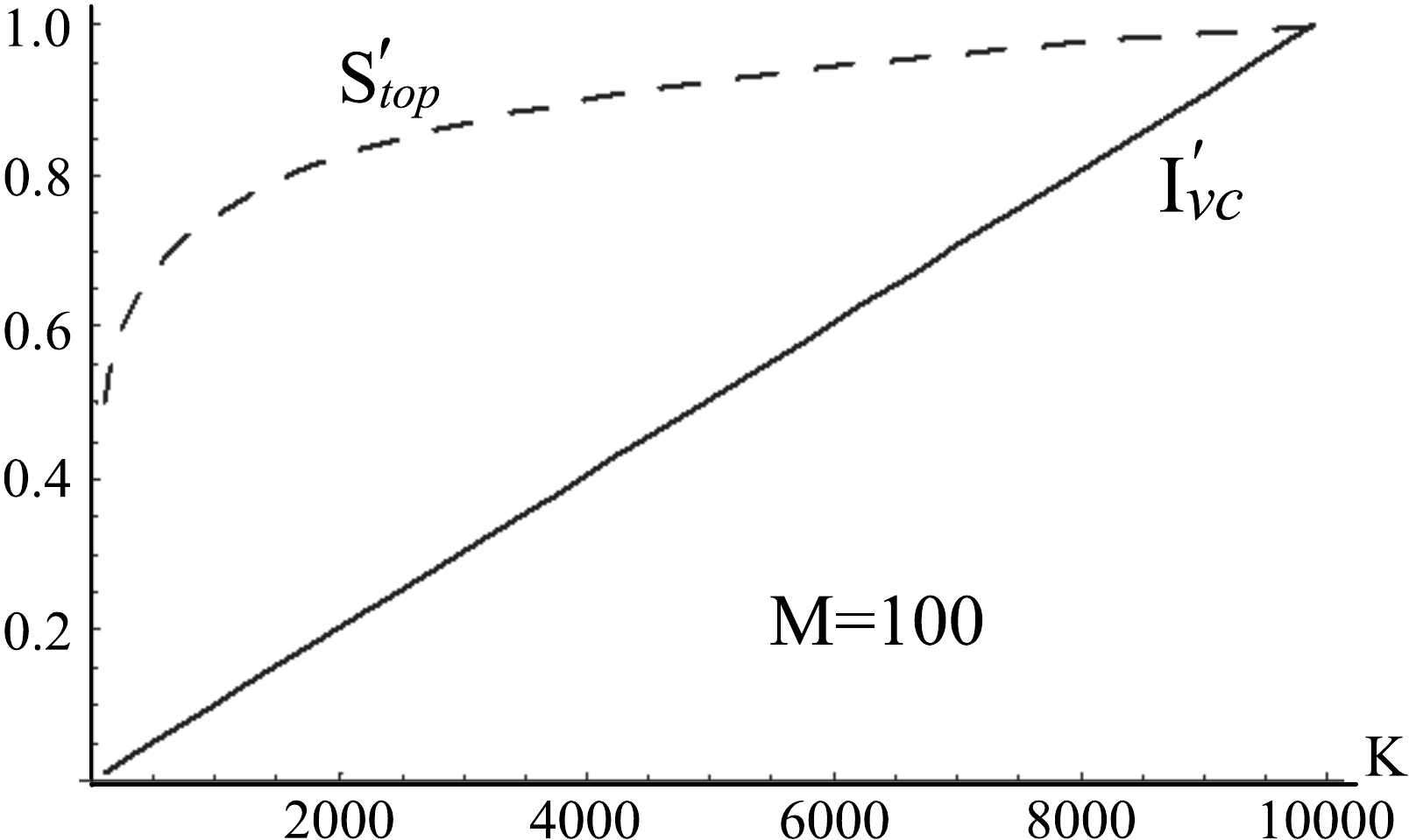}
\caption{The normalized upper bounds $S^\prime_{top}$ (truncated)
and $I^\prime_{vc}$ for 100-vertex networks as functions of the
total degree $K$. }\label{fig6}
\end{figure}

Although $S_{top}$ behaves like $I_{vc}$ for the five-vertex
networks in Fig. \ref{fig3}, they are different in a few aspects.
First, the topological entropy $S_{top}$ depends explicitly on the
total number of $M$ while $I_{vc}$ depends on $M$ implicitly. This
originates from the fact that the index $I_{vc}$ makes use of
Shannon's formula for the total information content of the vertex
distribution, while the topological entropy $S_{top}$ makes use of
the distribution of classical random walkers over nodes. Second,
the order of magnitude of $S_{top}$ for any $M$-vertex networks is
$O(M\log M)$. On the other hand, the order of magnitude of
$I_{vc}$ is $O(M\log M)$ for the network of a star-like topology,
and is $O(M^2 \log M)$ for the network of a fully connected
topology. It means that $I_{vc}$ increases much more rapidly than
$S_{top}$ when more and more edges are appearing between nodes.
This is shown in Fig. \ref{fig6} for 100-vertex networks. For any
networks with fixed number of nodes $M$ and fixed total degree
$K$, the indices $I_{vc}(M, K;\{K_a\})$ and $S_{top}(M,
K;\{K_a\})$ descriptors are viewed as functions of the node
distribution $\{K_a\}$. Theoretically, there are upper bounds for
$I_{vc}(M, K;\{K_a\})$ and $S_{top}(M, K;\{K_a\})$ with fixed
values of $M$ and $K$
\begin{eqnarray}
I_{vc}(M, K;\{K_a\}) &\leq& I_{vc}(M, K;\{K_a=K/M\}) = K \log_2 M, \label{Ivc3}\\
S_{top}(M, K;\{K_a\}) &\leq& S_{top}(M, K;\{K_a = K/M\}) \nonumber
\\
&=&\frac{M-1}{2}[\ln (2\pi)+1] -\frac{M}{2}\ln M + (M-1)\ln K.
\label{Stop3}
\end{eqnarray}
\noindent The results in the above equations show that the upper
bound of $I_{vc}$ increases linearly with $K$ for $M$-node
networks, and the upper bound of $S_{top}$ is a linear function of
$\log K$ for $M$-node networks. After truncating the
$K$-independent terms in (\ref{Stop3}), the truncated upper bound
of $S_{top}$ is normalized as
\begin{equation}
S^\prime_{top} = \frac{\ln K}{\ln (M(M-1))}. \label{Stop4}
\end{equation}
\noindent The normalized upper bound $I^\prime_{vc}$ is defined as
\begin{equation}
I^\prime_{vc} = \frac{K}{M(M-1)}. \label{Ivc4}
\end{equation}
\noindent Fig. \ref{fig6} shows the normalized upper bounds
$I^\prime_{vc}$ and $S^\prime_{top}$ for $M=100$. The total degree
$K$ ranges from $2(M-1)$ to $M(M-1)$ in Eq.(\ref{Ivc4}) and
(\ref{Stop4}). When $M$ is large, $S^\prime_{top}$ goes
approximately from $0.5$ to $1.0$ as $K$ increases, different to
the upper bound $I^\prime_{vc}$ that goes approximately from zero
to $1.0$. For most of the values of $K$, $S^\prime_{top}$ is a
slow-changing function of $K$ as compared to $I^\prime_{vc}$.

\begin{acknowledgments}
The author thanks Dr. Ding-Wei Huang for helpful discussions and
comments.
\end{acknowledgments}



\newpage


\end{document}